\documentclass[11pt]{article}
\usepackage{bbm}
\usepackage{mathrsfs}
\usepackage{amsfonts}
\usepackage{float}
\usepackage{amsthm}
\usepackage{epsfig,rotating}
\usepackage{amssymb,amsmath}
\usepackage{latexsym}
\usepackage{appendix}
\usepackage{graphicx,subfigure}
\usepackage{multirow}
\usepackage{color}
\usepackage{booktabs}
\usepackage[round, authoryear]{natbib}
\usepackage[ruled,linesnumbered]{algorithm2e}
\usepackage{subfigure}
\usepackage{natbib,float}
\usepackage{comment}
\usepackage{xcolor}

\newtheorem{theorem}{Theorem}

\newtheorem{proposition}{Proposition}

\def\bx{\boldsymbol{x}}
\def\bX{\boldsymbol{X}}

\def\bw{\boldsymbol{\omega}}

\def\bbeta{\boldsymbol{\beta}}
\def\bbetat{\widetilde{\boldsymbol{\beta}}}

\def\b1{\boldsymbol{1}}

\newcommand{\be}{\begin{equation}}
\newcommand{\ee}{\end{equation}}
\newcommand{\beaa}{\begin{eqnarray*}}
\newcommand{\eeaa}{\end{eqnarray*}}
\newcommand{\bea}{\begin{eqnarray}}
\newcommand{\eea}{\end{eqnarray}}
\newcommand{\bal}{\begin{align}}
\newcommand{\eal}{\end{align}}
\newcommand{\bali}{\begin{align*}}
\newcommand{\eali}{\end{align*}}

\providecommand{\keywords}[1]
{	
	\textbf{\textit{Keywords:}} #1
}

\title{
Statistical Inference for High-Dimensional Robust Linear Regression Models via Recursive Online-Score Estimation
}
\author{ Dian Zheng and Lingzhou Xue \\ Department of Statistics, The Pennsylvania State University
}

\date{}

\begin{document}
\maketitle
\begin{abstract}
    This paper introduces a novel framework for estimation and inference in penalized M-estimators applied to robust high-dimensional linear regression models. Traditional methods for high-dimensional statistical inference, which predominantly rely on convex likelihood-based approaches, struggle to address the nonconvexity inherent in penalized M-estimation with nonconvex objective functions. Our proposed method extends the recursive online score estimation (ROSE) framework of \cite{shi} to robust high-dimensional settings by developing a recursive score equation based on penalized M-estimation, explicitly addressing nonconvexity. We establish the statistical consistency and asymptotic normality of the resulting estimator, providing a rigorous foundation for valid inference in robust high-dimensional regression. The effectiveness of our method is demonstrated through simulation studies and a real-world application, showcasing its superior performance compared to existing approaches.
\end{abstract}


\section{Introduction}
In recent years, high-dimensional inference has garnered significant research interest, with a particular focus on developing robust statistical methods for analyzing high-dimensional data. One active area of research involves testing the mean or covariance matrix, as demonstrated in studies such as \cite{chen2010two,chen2010tests,cai2011limiting,cai2013two,tony2014two,li2015joint,li2018applications,li2024power,li2025power,cui2024hypothesis,shijie, zhe} and \cite{yu2023power,yu2024fisher,yu2024power,yu2025unified}. 

Another prominent direction explores inference for high-dimensional regression coefficients in linear and generalized linear models. For instance, \cite{zhang-zhang}  and \cite{Java} proposed using bias-corrected linear estimators derived from the Lasso to construct confidence intervals for regression coefficients. In addition, \cite{liu_yu} and \cite{liu_etal} developed inference techniques leveraging bootstrapping with modified least squares and partial ridge estimators, respectively. Extending these efforts, \cite{vande} adapted the de-sparsified Lasso estimator for generalized linear models, building on the foundation established by \cite{zhang-zhang}. Furthermore, \cite{ning_liu} introduced a decorrelated score statistic specifically designed for high-dimensional penalized M-estimators, providing an alternative approach for constructing confidence intervals. More recently, \cite{shi} proposed the recursive online-score estimation (ROSE) method, which combines model selection with the recursive construction of score equations.


While the existing literature has made significant progress in high-dimensional inference under convex settings, many practical challenges remain unaddressed. In particular, these convex approaches often assume light-tailed errors and clean data, which may not hold in real-world applications.
In contrast, scenarios involving heavy-tailed errors, adversarial contamination, or complex noise structures require robust regression techniques that naturally lead to nonconvex optimization problems. Such nonconvexity is not merely a technical complication but an essential feature of models designed to reflect more realistic data-generating processes.

Consider the following dataset as an example. Riboflavin (vitamin B2) is a water-soluble nutrient produced by all plants and most microorganisms, essential for the growth and reproduction of humans and animals. The riboflavin production dataset, derived from \emph{Bacillus subtilis} and provided by DSM Nutritional Products, is available in the R package ``hdi". This dataset includes 71 observations, with a response variable representing the logarithm of the riboflavin production rate and 4088 explanatory variables capturing the logarithm of the expression levels of 4088 genes. The primary objective is to identify significant genes influencing riboflavin production. A key challenge in analyzing this dataset lies in the high dimensionality of predictors relative to the limited sample size. Furthermore, as noted by \cite{arashi}, the dataset contains certain outliers. To explore this issue, we initially modeled the data using a linear regression approach with the  {smoothly clipped absolute deviation (SCAD)}  penalty \citep{scad}. The residuals' skewness was assessed using both a histogram and a QQ plot, as illustrated below. Additionally, Mardia's skewness test \citep{mardia} rejected the null hypothesis of zero skewness (p-value = 0.0199), confirming the presence of significant skewness.

\begin{figure}[h]
    \centering
    \subfigure[The histogram and density plot of the residuals]{\label{Ch3-figure: hist}
    \includegraphics[width=0.45\linewidth]{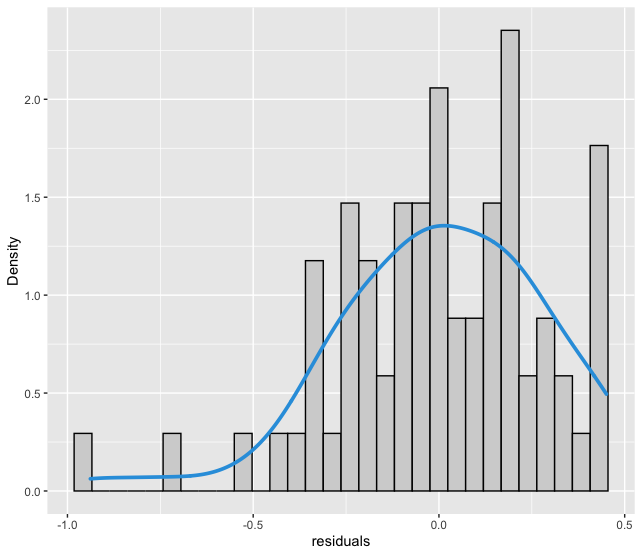}}
    \hfill
    \subfigure[The QQ plot of the residuals]{\label{Ch3-figure: qqplot}
    \includegraphics[width=0.45\linewidth]{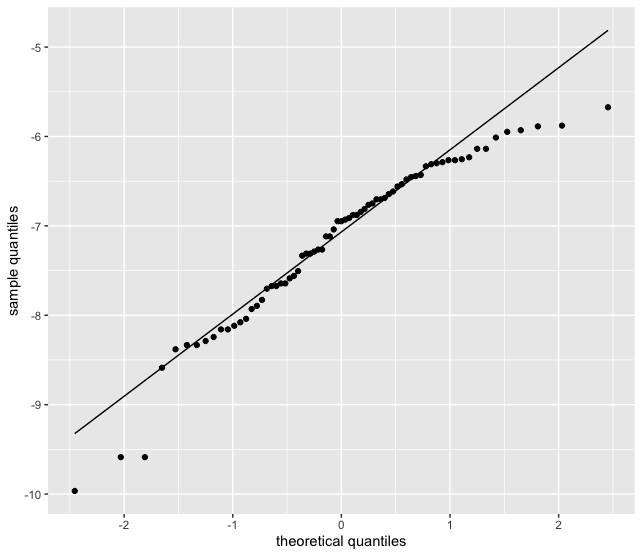}}
   
\end{figure}

Thus, the dataset poses two primary challenges: the high dimensionality of predictors and the presence of outliers and skewness in the data. While existing methods can address the first challenge, they are ill-equipped to directly tackle the second. This underscores the importance of nonconvex analysis in addressing these complex issues effectively.

Recent years have witnessed substantial progress in understanding high-dimensional penalized M-estimation problems with nonconvex objective functions. \cite{po15} established statistical bounds on the distance between any local optimum of the empirical objective and the unique minimizer of the population risk, provided the restricted strong convexity (RSC) condition holds. This pioneering work significantly advanced the literature by elucidating the tradeoff between statistical accuracy and optimization efficiency in high-dimensional settings. Building on this foundation, \cite{poling} examined the local behavior of stationary points near the true parameter vector in high-dimensional robust M-estimation under the local RSC condition. They demonstrated statistical consistency within the local region. Moreover, by employing suitable folded-concave regularizers such as SCAD \citep{scad} and MCP \citep{zhang2010}, \cite{zou2008one} and \cite{fan2014strong} showed that local stationary points of the regularized robust M-estimator exhibit the oracle property. A computational guarantee was achieved through a novel two-step procedure: with a convex M-estimator being used as the initial solution, a folded-concave M-estimator can achieve the strong oracle property.

In a related vein, \cite{Song} studied the function landscape and derived statistical error bounds for stationary points in high-dimensional M-estimation with nonconvex losses. Their conditions, while similar to RSC, are more straightforward to verify, broadening the applicability of their results. Following the philosophy of \cite{Song}, \cite{liu2024robust} proved the fine landscape of their nonconvex estimation procedure for robust high-dimensional regression with coefficient thresholding, and \cite{zheng2024smoothed} proved the benign landscape of smoothed robust phase retrieval. However, these approaches primarily focus on quantifying the distance between the estimators and the true parameters, addressing the problem from an estimation perspective rather than an inferential one. Consequently, they are not directly equipped to resolve the second challenge encountered in real data analysis, such as addressing skewness and outlier effects.

In this paper, we address both challenges simultaneously, contributing to the field by developing a recursive online estimation method for high-dimensional robust linear regression. This approach generalizes maximum likelihood estimation to accommodate nonconvexity and robustness. Specifically, we consider the high-dimensional linear model:
\be \label{model}
Y= \bX^T \boldsymbol{\beta}+\varepsilon,\ee
where $\boldsymbol{\beta}=\left(\beta_{1}, \beta_{2}, \ldots, \beta_{ p}\right)^T \in \mathbb{R}^p$ is the parameter vector of interest. Besides, $\varepsilon$ represents the potentially contaminated or heavy-tailed noise \citep{fan2016multitask,poling,Song,sun2020adaptive,li2023robustb,li2023robusta}. Our primary focus is on constructing confidence intervals (CIs) for a univariate parameter of interest, $\beta_{ j_0}$, for some $j_0 \in\{1, \ldots, p\}$. Motivated by the analysis of the riboflavin dataset, the central challenges in this problem include managing the high-dimensional nuisance parameters and addressing the nonconvexity inherent in the M-estimation process. By tackling these issues, our method provides a robust and efficient solution for inference in high-dimensional settings, offering both theoretical guarantees and practical applicability. 

Our approach begins with an in-depth examination of the penalized M-estimation landscape, building on the foundational work of \cite{Song}. We demonstrate that our proposed penalized M-estimation framework possesses a well-behaved optimization landscape with high probability. This ensures that the initial estimator for the entire parameter vector, $\bbeta$, exhibits desirable statistical properties, providing a robust foundation for subsequent inference. To facilitate practical implementation, we develop a composite gradient algorithm and establish its computational guarantees. Building on this initial estimator, we apply a recursive feature screening procedure to identify relevant variables and construct estimating equations based on the selected features. The final estimate is obtained by solving the online estimation equation, and we derive asymptotic results to characterize its statistical properties. Our method stands apart from the ROSE framework proposed by \cite{shi}, as it operates independently of likelihood assumptions. This makes it particularly well-suited for more challenging settings, such as high-dimensional robust regressions involving contaminated or heavy-tailed errors. In the following sections, we present simulation studies demonstrating the superior performance of our method—particularly in linear models subject to these complexities—when compared to ROSE and other benchmark methods.

 {
Our main contributions are summarized as follows. We propose a recursive online score estimation (ROSE) methodology tailored for high-dimensional robust linear regression, extending the original ROSE framework to more general and challenging nonconvex settings. To handle the inherent nonconvexity introduced by robust loss functions, we provide a careful analysis of the M-estimation landscape and establish theoretical guarantees for the proposed algorithm. Building upon an initial estimator, our method achieves asymptotic normality and enables valid statistical inference for low-dimensional components of high-dimensional parameters. Through comprehensive simulation studies and real data analysis, we demonstrate the practical advantages of our approach over existing methods, especially in the presence of heavy-tailed errors and data contamination. }

The rest of this paper is organized as follows: Section 2.1 introduces our methods, Section 2.2 discusses the landscape analysis, and Section 2.3 establishes asymptotic normality. In Section 3, we evaluate the performance of our proposed procedure through simulation studies, while in Section 4, we apply the method to a real dataset. Section 5 concludes the paper with a summary. The proofs are provided in the appendix  of the supplementary file.

\section{Methodology}

Before proceeding to our main method, some notations are introduced. We use bold capitalized italic letters (e.g., $\boldsymbol{X}, \bbeta$) and bold  letters in Roman font (e.g., $\boldsymbol{\mathrm{X}}, \boldsymbol{\mathrm{A}}$) to denote vectors and matrices repectively, and use  regular letters (e.g., $X, y$) to denote scalars. For a $p$-dimensional vector $\bX=\left(X_1, \ldots, X_p\right)^T$, we denote its $\ell_{q^{-}}$norm as $\|\bX\|_q=\left(\sum_{i=1}^p\left|X_i\right|^q\right)^{1 / q}$ $(q \in(0,2])$, and $\ell_0$ ``norm'' as $\|\boldsymbol{X}\|_0=\#\left\{j: X_j \neq 0\right\}$. For a matrix $\boldsymbol{\mathrm{A}}_{p \times q}=\left[a_{i j}\right]_{p \times q}$, its 1-norm, 2-norm, $\infty$-norm, and max-norm are defined as $\|\boldsymbol{\mathrm{A}}\|_1=\sup _j \sum_{i=1}^p\left|a_{i j}\right|,\|\boldsymbol{\mathrm{A}}\|_2=\max _{\boldsymbol{x}:\|\boldsymbol{x}\|_2=1}\|\boldsymbol{\mathrm{A} x}\|_2,\|\boldsymbol{\mathrm{A}}\|_{\infty}=$ $\sup _i \sum_{j=1}^q\left|a_{i j}\right|$, and $\|\boldsymbol{\mathrm{A}}\|_{\max }=\sup _{i, j}\left|a_{i j}\right|$, respectively. 
The $\ell_p$ norm of a vector $\bx$ is indicated by $\|\bx\|_p$. 
We let $\mathrm{B}_q^d(\mathbf{a}, \rho) \equiv\left\{\bx \in \mathbb{R}^d:\|\bx-\boldsymbol{a}\|_q \leq \rho\right\}$ be the $\ell_q$ ball in $\mathbb{R}^d$ with center $\mathbf{a}$  and radius $\rho$. We will often omit the dimension superscript $d$ when clear from the context, the subscript $q$ when $q=2$, and the center $\mathbf{a}$ when $\mathbf{a}=\mathbf{0}$. In particular, $\mathrm{B}(\rho)$ is the Euclidean ball of radius $\rho$. 

Denote the true value of $\boldsymbol{\beta}$ as $\boldsymbol{\beta}_0=\left(\beta_{0,1}, \ldots, \beta_{0, p}\right)^T$. Let $\boldsymbol{\Sigma}=\mathrm{E} \left[\frac{\partial^2}{\partial \boldsymbol{\beta} \partial \boldsymbol{\beta}^T} \ell(Y-\bX^T\bbeta)\right]$. For any $r \times q$ matrix $\boldsymbol{\Phi}$ and any sets $J_1 \subseteq[1, \ldots, r], J_2 \subseteq[1, \ldots, q]$, we denote by $\boldsymbol{\Phi}_{J_1, J_2}$ the submatrix of $\boldsymbol{\Phi}$ formed by rows in $J_1$ and columns in $J_2$. Similarly, for any $q$-dimensional vector $\boldsymbol{\psi}$, $\boldsymbol{\psi}_{J_1}$ stands for the subvector of $\boldsymbol{\psi}$ formed by elements in $J_1$. Let $\left|J_1\right|$ be the number of elements in $J_1$. Denote by $\mathcal{M}_{j_0}=\{j \neq$ $\left.j_0: \beta_{0, j} \neq 0\right\}$. Let $\mathbb{I}=\{1, \ldots, p\}$ and $\mathbb{I}_{j_0}=\mathbb{I}-\left\{j_0\right\}$. For any set $\mathcal{M} \subseteq \mathbb{I}_{j_0}$, define $\boldsymbol{\omega}_{\mathcal{M}, j_0}=\boldsymbol{\Sigma}_{\mathcal{M}, \mathcal{M}}^{-1} \boldsymbol{\Sigma}_{\mathcal{M}, j_0}$ and
$$
\sigma_{\mathcal{M}, j_0}^2=\mathrm{E}\left(\frac{\partial \ell\left(Y-\bX^T\bbeta\right)}{\partial \beta_{j_0}}-\boldsymbol{\omega}_{\mathcal{M}, j_0}^T \frac{\partial \ell\left(Y-\bX^T\bbeta\right)}{\partial \boldsymbol{\beta}_{\mathcal{M}}}\right)^2.
$$

Let $\|\boldsymbol{Z}\|_{\psi_p}$ be the Orlicz norm of any random variable $\boldsymbol{Z}$,
$$
\|\boldsymbol{Z}\|_{\psi_p} \triangleq \inf _{c>0}\left\{\mathrm{E} \exp \left(\frac{|\boldsymbol{Z}|^p}{c^p}\right) \leq 2\right\}.
$$

 {Let $\|f\|_{\infty}$ be the infinity norm of  any function $f\in L^\infty(\mathbb{R}),$
$$
\|f\|_{\infty} \triangleq\inf \{C \geq 0:|f(x)| \leq C \text { for almost every } x \in \mathbb{R}\} .
$$}

\subsection{Estimation and inference}

Suppose that $\left(\boldsymbol{X}_1, Y_1\right), \ldots,\left(\boldsymbol{X}_n, Y_n\right)$ is a random sample from the model (\ref{model}). The dimension $p$ satisfies $\log p = o(n)$.
 We aim to construct a CI for $\beta_{j_0},$ the parameter of interest in the presence of the complement $\bbeta_{\mathbb{I}_{j_0}}$, which is a high-dimensional nuisance vector. To deal with it, \cite{ning_liu} proposed the following decorrelated score function
 \begin{equation}\label{descore}
\hat{S}(\beta_{j_0}, \widetilde{\bbeta}_{\mathbb{I}_{j_0}})=\sum_{t=1}^{n}\left(\frac{\partial \ell(Y_t-X_{t,j0}\beta_{j_0}- {\bX^T_{t,\mathbb{I}_{j_0}}\bbetat_{\mathbb{I}_{j_0}}})}{\partial \beta_{j_0}}-\hat{\bw}^T \frac{\partial \ell(Y_t-X_{t,j0}\beta_{j_0}- {\bX^T_{t,\mathbb{I}_{j_0}}\bbetat_{\mathbb{I}_{j_0}}})}{\partial \bbeta_{\mathbb{I}_{j_0}}}\right),
\end{equation}
where  {$\ell$ is a general loss function},  $\bbetat$ is an initial estimator for the whole vector and $\hat{\bw}$ is Dantzig type estimator of $\bw_{\mathbb{I}_{j_0},j_0}$.

To improve estimation efficiency and address the issue of high dimensionality, we leverage the idea of recursive online learning introduced by \cite{shi}.
Specifically, we consider the following score function instead:

\begin{equation}\label{rose}
\sum_{t=0}^{n-1}\left(\frac{\partial \ell(Y_{t+1}-X_{t+1,j0}\beta_{j_0}- {\bX^T_{t+1,\widehat{\mathcal{M}}_{j_0}^{(t)}}\bbetat_{\widehat{\mathcal{M}}_{j_0}^{(t)}})}}{\partial \beta_{j_0}}-\hat{\bw}_{\widehat{\mathcal{M}}_{j_0}^{(t)}, j_0}^T \frac{\partial \ell(Y_{t+1}-X_{t+1,j0}\beta_{j_0}- {\bX^T_{t+1,\widehat{\mathcal{M}}_{j_0}^{(t)}}\bbetat_{\widehat{\mathcal{M}}_{j_0}^{(t)}})}}{\partial \bbeta_{\widehat{\mathcal{M}}_{j_0}^{(t)}}}\right).  
\end{equation}
Here, $\bbetat$ is still an initial estimator, derived from 
$$\underset{\boldsymbol{\beta} \in\mathrm{B}(r)}{\min }\left(\frac{1}{n} \sum_{i=1}^n \ell\left( Y_i-\bX_i^T\bbeta\right)+\sum_{j=1}^p  {\lambda}\left|\beta_j\right|\right),$$
where  $\mathrm{B}(r)$ is the $p$-dimensional ball centered at origin with radius $r$. The radius 
$r$
 should be chosen sufficiently large to ensure that the true parameter vector 
$\bbeta_0$ is feasible.  {The function $l(\cdot)$ is a potentially nonconvex loss function,  often either bounded or increasing slowly at infinity, as naturally suggested by robustness considerations \citep{fan2016multitask,poling,Song,sun2020adaptive,li2023robustb,li2023robusta}. The classical choices of $l(\cdot)$ include the Huber loss and Tukey's bisquare loss. }
This corresponds to Step 2 in our proposed procedure below. In the next subsection, we will elaborate on the theoretical properties of 
$\bbetat$ via solving this possibly nonconvex M-estimation program with a proposed algorithm. 

The key difference between \eqref{descore} and \eqref{rose} is the data-splitting strategy for model selection and score evaluation in the latter. Figure \ref{fig-rose} below showcases the overview of the methodology. Specifically,  for each $t$,  we use the sub-dataset
$\mathcal{F}_t=$ $\left\{\left(\boldsymbol{X}_1, Y_1\right), \ldots,\left(\boldsymbol{X}_t, Y_t\right)\right\}$ to estimate the support set of $\bbeta$ except $\beta_{j_0}$. The resulting estimator is denoted as  $\widehat{\mathcal{M}}_{j_0}^{(t)}$. In order to ensure the effectiveness of model selection, $\mathcal{F}_t$ cannot be too small. Similar to the approach in \cite{shi}, when $t>s_n$, where $s_n$ is a prespecified integer, we estimate $\mathcal{M}_{j_0}$ based on 
 $\mathcal{F}_t$  while when $t\le s_n$, the estimator 
$\widehat{\mathcal{M}}_{j_0}^{(t)}$ is calculated based on the sub-dataset $\left\{\left(\boldsymbol{X}_{s_n+1}, Y_{s_n+1}\right), \ldots,\left(\boldsymbol{X}_n, Y_n\right)\right\}$. In this way, $\mathbb{I}_{j_0}$ in \eqref{descore} could be replaced by $\widehat{\mathcal{M}}_{j_0}^{(t)}$, which significantly reduces the dimension of nuisance parameter. Therefore, instead of using the Dantzig-type estimator mentioned earlier, we employ a plug-in estimator for $\bw$. Besides, given $\widehat{\mathcal{M}}_{j_0}^{(t)}$, we need to calculate the estimator of $\sigma_{\mathcal{M}, j_0}^2$, which is necessary for establishing  the asymptotic normality result. 

\begin{figure}[H]
    \centering
    \includegraphics[width=9cm]{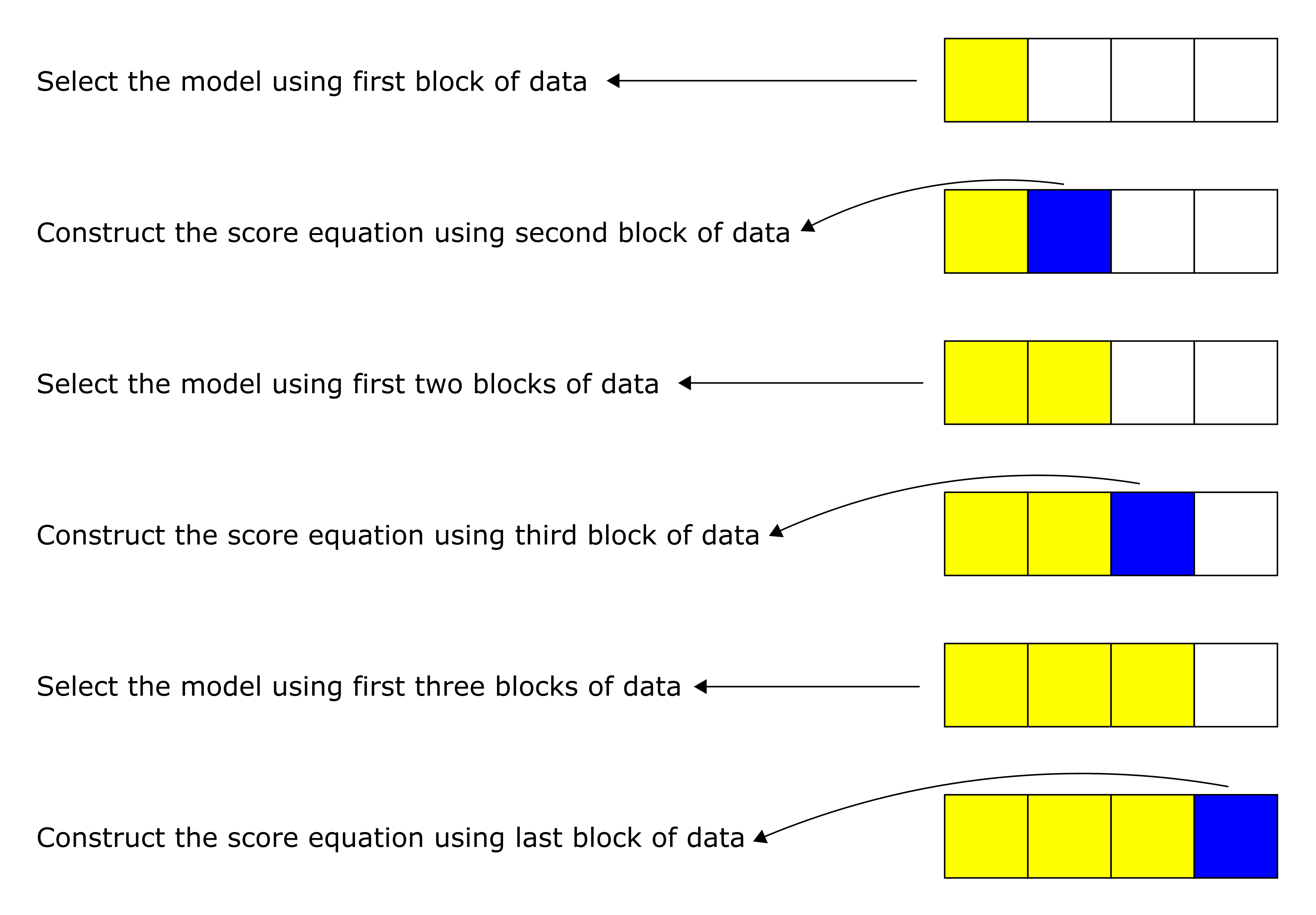}
    \caption{Overview of the ROSE method}
    \label{fig-rose}
\end{figure}

The estimating procedures are as follows.

\textbf{Step 1.} Input $\left\{\boldsymbol{X}_i, Y_i\right\}_{i=1}^n$ and an integer $1<s_n<n$.\par
\textbf{Step 2.} Compute an initial estimator $\widetilde{\boldsymbol{\beta}}$ for $\boldsymbol{\beta}_0$. Compute
$$
\widehat{\boldsymbol{\Sigma}}=\frac{1}{n} \sum_{i=1}^n \frac{\partial^2}{\partial \boldsymbol{\beta} \partial \boldsymbol{\beta}^T} \ell\left(Y_i-\bX_i^T\bbetat\right).
$$ \par
\textbf{Step 3.}  For $t=s_n, s_n+1, \ldots, n-1$, estimate $\mathcal{M}_{j_0}$ based on the sub-dataset $\mathcal{F}_t=$ $\left\{\left(\boldsymbol{X}_1, Y_1\right), \ldots,\left(\boldsymbol{X}_t, Y_t\right)\right\}$. Denoted by $\widehat{\mathcal{M}}_{j_0}^{(t)}$ the corresponding estimator. We require $\left|\widehat{\mathcal{M}}_{j_0}^{(t)}\right| \leq n, j_0 \notin \widehat{\mathcal{M}}_{j_0}^{(t)}$. Compute
$$
\begin{aligned}
&\widehat{\boldsymbol{\omega}}_{\widehat{\mathcal{M}}_{j_0}^{(t)}, j_0}=\widehat{\boldsymbol{\Sigma}}^{-1}_{\widehat{\mathcal{M}}_{j_0}^{(t)}, \widehat{\mathcal{M}}_{j_0}^{(t)}} \widehat{\boldsymbol{\Sigma}}_{\widehat{\mathcal{M}}_{j_0}^{(t)}, j_0}, \quad \text{and}\\
&\hat{\sigma}_{\widehat{\mathcal{M}}_{j_0}^{(t)}, j_0}^2=\frac{1}{n} \sum_{i=1}^n 
\left(\frac{\partial \ell(Y_i-\bX_i^T\bbetat)}{\partial \beta_{j_0}}-\widehat{\boldsymbol{\omega}}_{\widehat{\mathcal{M}}_{j_0}^{(t)}, j_0}^T \frac{\partial \ell(Y_i-\bX_i^T\bbetat)}{\partial \boldsymbol{\beta}_{\widehat{\mathcal{M}}_{j_0}^{(t)}, j_0}}\right)^2.
\end{aligned}
$$\par
\textbf{Step 4.}  Estimate $\mathcal{M}_{j_0}$ based on the sub-dataset $\left\{\left(\boldsymbol{X}_{s_n+1}, Y_{s_n+1}\right), \ldots,\left(\boldsymbol{X}_n, Y_n\right)\right\}$. Denoted by $\widehat{\mathcal{M}}_{j_0}^{\left(-s_n\right)}$ the resulting estimator. We require $\left|\widehat{\mathcal{M}}_{j_0}^{\left(-s_n\right)}\right| \leq n, j_0 \notin \widehat{\mathcal{M}}_{j_0}^{\left(-s_n\right)}$. Compute
$$
\begin{aligned}
&\widehat{\boldsymbol{\omega}}_{\widehat{\mathcal{M}}_{j_0}^{\left(-s_n\right)}, j_0}=\widehat{\boldsymbol{\Sigma}}^{-1}_{\widehat{\mathcal{M}}_{j_0}^{\left(-s_n\right)}, \widehat{\mathcal{M}}_{j_0}^{\left(-s_n\right)}} \widehat{\boldsymbol{\Sigma}}_{\widehat{\mathcal{M}}_{j_0}^{\left(-s_n\right)}, j_0} \text { and }\\
&\hat{\sigma}_{\widehat{\mathcal{M}}_{j_0}^{(-s_n)}, j_0}^2=\frac{1}{n} \sum_{i=1}^n 
\left(\frac{\partial \ell(Y_i-\bX_i^T\bbetat)}{\partial \beta_{j_0}}-\widehat{\boldsymbol{\omega}}_{\widehat{\mathcal{M}}_{j_0}^{\left(-s_n\right)}, j_0} ^T \frac{\partial \ell(Y_i-\bX_i^T\bbetat)}{\partial \boldsymbol{\beta}_{\widehat{\mathcal{M}}_{j_0}^{(-s_n)}, j_0}}\right)^2.
\end{aligned}
$$

\textbf{Step 5.}  Define $\hat{\beta}_{j_0}$ to be the solution to the following equation,
$$
\begin{aligned}
& \sum_{t=0}^{s_n-1} \frac{\widehat{Z}_{t+1, j_0}}{\hat{\sigma}_{\widehat{\mathcal{M}}_{j_0}^{\left(-s_n\right)}, j_0}}\left\{\ell'\left(Y_{t+1}-(X_{t+1, j_0} \beta_{ j_0}+\boldsymbol{X}_{t+1, \widehat{\mathcal{M}}_{j_0}^{\left(-s_n\right)}} \widetilde{\boldsymbol{\beta}}_{\widehat{\mathcal{M}}_{j_0}^{\left(-s_n\right)}})\right)\right\} \\
& +\sum_{t=s_n}^{n-1} \frac{\widehat{Z}_{t+1, j_0}}{\hat{\sigma}_{\widehat{\mathcal{M}}_{j_0}^{(t)}, j_0}}\left\{\ell'\left(Y_{t+1}-(X_{t+1, j_0} \beta_{ j_0}+\boldsymbol{X}_{t+1, \widehat{\mathcal{M}}_{j_0}^{\left(t\right)}} \widetilde{\boldsymbol{\beta}}_{\widehat{\mathcal{M}}_{j_0}^{\left(t\right)}})\right)\right\} \\
&=0,
\end{aligned}
$$
The estimating equation in Step 5 can be solved via the Newton-Raphson method with the initial value $\hat{\beta}_{j_0}^{(0)}=\widetilde{\beta}_{j_0}$. More specifically, for $l=1,2, \ldots$, we can iteratively update $\hat{\beta}_{j_0}$ by

$$
\hat{\beta}_{j_0}^{(l)}=\hat{\beta}_{j_0}^{(l-1)}-\frac{\sum_{t=0}^{n-1} \frac{1}{\hat{\sigma}_{\widehat{\mathcal{M}}_{j_0}^{(t)}, j_0}} \widehat{Z}_{t+1, j_0}\left\{\ell'\left(Y_{t+1}-(X_{t+1, j_0} \beta_{ j_0}+\boldsymbol{X}_{t+1, \widehat{\mathcal{M}}_{j_0}^{\left(t\right)}} \widetilde{\boldsymbol{\beta}}_{\widehat{\mathcal{M}}_{j_0}^{\left(t\right)}})\right)\right\}}{\underbrace{\sum_{t=0}^{n-1} \frac{1}{\hat{\sigma}_{\widehat{\mathcal{M}}_{j_0}^{(t)}, j_0}} \widehat{Z}_{t+1, j_0} \dfrac{\partial}{\partial \beta_{j0}} \left\{\ell'\left(Y_{t+1}-(X_{t+1, j_0} \beta_{ j_0}+\boldsymbol{X}_{t+1, \widehat{\mathcal{M}}_{j_0}^{\left(t\right)}} \widetilde{\boldsymbol{\beta}}_{\widehat{\mathcal{M}}_{j_0}^{\left(t\right)}})\right)\right\} }_{\Gamma_n^{*,(l-1)}}},
$$
where we use a shorthand and write $\widehat{\mathcal{M}}_{j_0}^{(t)}=\widehat{\mathcal{M}}_{j_0}^{\left(-s_n\right)}$, for $t=0, \ldots, s_n-1$. \par
A two-sided $1-\alpha$ CI for $\beta_{ j_0}$ is given by
$$
\hat{\beta}_{j_0}^{(l)} \pm \frac{z_{\frac{\alpha}{2}}}{\sqrt{n} \Gamma_n^{*,(l-1)}}.
$$

Our method builds on the same core idea of recursive online learning as in ROSE \citep{shi}, and similarly derives a recursive form for constructing confidence intervals. However, our generalization to the robust high-dimensional setting is far from straightforward. As will be discussed in the next section, this extension requires a careful investigation of the optimization landscape under nonconvex loss functions, along with the establishment of key theoretical properties. These elements are essential to ensure the validity of the proposed inference procedure and distinguish our method from the original ROSE framework.

\subsection{Initial estimator}

\subsubsection{Landscape analysis}

Before establishing the asymptotic validity of the  proposed CIs, we first study the landscape of the following problem involved in step 2:
\begin{equation}\label{ini}
\begin{array}{ll}
\operatorname{minimize} & \frac{1}{n} \sum_{i=1}^n\ell\left(Y_i-\bX_i^T\bbeta\right)+\lambda_n\|\boldsymbol{\beta}\|_1, \\
\text { subject to } & \|\boldsymbol{\beta}\|_2 \leq r .
\end{array}
\end{equation}

We state the following assumptions. \par

\textbf{Assumption 1.}\par
(a) The derivative of loss, i.e. $\ell'$, satisfies $\sup _{t \in \mathbb{R}}\{|\ell'(t) t|\} \leq$ $C_\ell$ for some absolute constant $C_\ell$. \par

(b) $\ell'$ is odd, $\ell'(z)\ge0, \forall z\ge0$ and $\exists L>0, \max\{\|\ell'(Y-\bX^T\bbeta)\|_\infty,\|\ell''(Y-\bX^T\bbeta)\|_\infty,\|\ell'''(Y-\bX^T\bbeta)\|_\infty\}\le L.$\par

(c) The feature vector $\bX$  is bounded: $\|\bX\|_{\infty} \leq M \tau$, and $\left|\left\langle\bX, \bbeta_0 /\left\|\bbeta_0\right\|_2\right\rangle\right| \leq M \tau$ almost surely, with $\bbeta_0$ the ground truth parameter. Here, $M$ is a dimensionless constant greater than 1. 

\par 

(d) The feature vector $\bX$  spans all directions in $\mathbb{R}^d$, that is, $\mathbb{E}\left[\bX \bX^{\top}\right] \succeq \underline{\gamma}M^2 \tau^2 \mathbf{I}_{d \times d}$ for some $0<\underline{\gamma}<1$.

\par 

(e)  The noise $\varepsilon$ has a symmetric distribution. Further, defining $h(z) \equiv \mathbb{E}_{\varepsilon}\{\ell'(z+\varepsilon)\}$ we have $h(z)>0$ for all $z>0$, as well as $h^{\prime}(0)>0$.\par 

\vspace{10pt}

Condition (a)-(d) present technical conditions on the loss and design matrix respectively. These are standard conditions in nonconvex landscape analysis, as also outlined in \cite{Song}.
 {Condition (a) imposes a fast-decaying assumption on the derivative of loss $\ell$. Condition (b) further assumes  that the  derivatives of $\ell$ up to third order are bounded over the range of ($Y-\bX^T\bbeta$). Condition (c) requires boundness of the covariate $\bX$ as well as their inner product with the standardized true parameter. while condition (d) assumes the covariance matrix of $\bX$ is positive definite. }
As pointed out by \cite{Song}, condition (e) is relatively mild and can be satisfied, for instance, if the noise has a density that is strictly positive and decreases when $\varepsilon>0.$

\par

\vspace{10pt}
\begin{theorem}
Under Assumption 1, further assume $\left\|\boldsymbol{\beta}_0\right\|_0 \leq s_0$ and $\left\|\boldsymbol{\beta}_0\right\|_2 \leq r / 2$. Then there exist constants $C_n, C_\lambda, C_s$ and $\varepsilon_0$ depending on $\left(L, C_g, r, \tau^2, \underline{\gamma}, \delta\right)$ and the function $g(\cdot)$, but independent of $n$, $p$,  $s_0$ and $M$, such that as $n \geq C_n s_0 \log p$ and $\lambda_n \geq C_\lambda M \sqrt{(\log p) / n}$, the following hold with probability at least $1-\delta$ : \par
(a) Any stationary point $\check\bbeta$ of problem \eqref{ini} is in $\mathrm{B}_2^d\left(\boldsymbol{\theta}_0, C_s\left(\left(M^2 s_0 \log d\right) / n+s_0 \lambda_n^2\right)^{1 / 2}\right)$ and satisfies
$\left\|\check{\boldsymbol{\beta}}_{\mathcal{M}_0^c}-\boldsymbol{\beta}_{0, \mathcal{M}_0^c}\right\|_1 \leq 3\left\|\check{\boldsymbol{\beta}}_{\mathcal{M}_0}-\boldsymbol{\beta}_{0, \mathcal{M}_0}\right\|_1$.
\par
(b) As long as $n$ is large enough such that $n \geq C_n s_0 \log ^2 p$ and $C_s\left(\left(M^2 s_0 \log p\right) /\right.$ $\left.n+s_0 \lambda_n^2\right)^{1 / 2} \leq \varepsilon_0$, and the feature vector $\bX$ is continuous, the problem has a unique local minimizer $\widehat{\boldsymbol{\beta}}_n$ which is also the global minimizer.
\end{theorem}

Theorem 1 provides $\ell_1 / \ell_2$ estimation error bounds for all the stationary points, which is crucial for further inference. 
 Moreover, part (b) proves the uniqueness of the solution if $n$ meets a stricter order criterion, in addition to the continuity requirement being fulfilled.
The result in Part (a) is sufficient for our theoretical analysis and Theorem 1 aligns with the findings in \cite{Song}.

\subsubsection{Computational guarantee}

We propose to use the composite gradient descent algorithm  {\citep{nest}}, which is computationally efficient for solving the nonconvex optimization and enjoys the convergence property. \par

Consider the  empirical risk function 
$$\widehat{R}_n\left(\bbeta\right)\triangleq \frac{1}{n} \sum_{i=1}^n\ell\left(Y_i-\bX_i^T\bbeta\right), $$

 we can rewrite the program  {(2)} as
$$
\widehat{\bbeta} \in \arg \min _{\|\bbeta\|_2 \leq r}\left\{\widehat{R}_n(\bbeta)+\lambda\|\bbeta\|_1\right\}.
$$

Specifically, it contains two key steps at each iteration: the gradient descent step and the $\ell_2$-ball projection step.

In the first step, we perform gradient descent. Given the previous iterated solution $\hat{\beta}^{(k)}$, with the step size $h$, we need to solve the following subproblem:
\begin{equation}
\min_{\bbeta}\left\{\frac{1}{2}\left\|\bbeta-\left(\hat{\bbeta}^{(k)}-\frac{1}{h} \nabla \hat{R}_n\left(\hat{\bbeta}^{(k)}\right)\right)\right\|_2^2+\frac{\lambda}{h} \sum_{j=1}^p\left\|\beta^j\right\|_1\right\} .
\end{equation}

Defining the soft-thresholding operator $S_{\lambda / \eta}(\beta)$ componentwise according to
$$
S_{\lambda / \eta}^j:=\operatorname{sign}\left(\beta_j\right)\left(\left|\beta_j\right|-\frac{\lambda}{\eta}\right)_{+}.
$$

 Thus, the gradient descent step can be solved as
$$
\tilde{\bbeta}^{(k+1)}=S_{\lambda / h}\left(\hat{\bbeta}^{(k)}-h \nabla\left(\widehat{R}_n\left(\hat{\bbeta}^{(k)}\right)\right)\right) .
$$
In the second step, we project $\tilde{\bbeta}^{(k+1)}$ onto the $\ell_2$-ball by
$$
\pi_r\left(\tilde{\bbeta}^{(k+1)}\right)=\frac{\min \left\{\left\|\tilde{\bbeta}^{(k+1)}\right\|_2, r\right\}}{\left\|\tilde{\bbeta}^{(k+1)}\right\|_2} \tilde{\bbeta}^{(k+1)}.
$$

\par

\vspace{10pt}
The proposed algorithm can be summarized as Algorithm 1.

\par
\vspace{10pt}
\begin{algorithm}[H]
	\caption{The composite gradient descent algorithm}
	\small
	\KwIn{$\bbeta^{(0)} \in B^p(r)$, step size $h$, penalization parameter $\lambda$, thresholding parameter $\eta$, \\ and predetermined hyperparameter $\tau, r$  }
	\KwOut{ $\widehat{\bbeta}$}
	
	\For{$k=0,1,2,\ldots$  until convergence }
	{
		$\begin{aligned} & \widetilde{\bbeta}^{(k+1)}=S_{\lambda / h}\left(\widehat{\bbeta}^{(k)}-h \nabla\left(\widehat{R}_n\left(\widehat{\bbeta}^{(k)}\right)\right)\right) \\ & \widehat{\bbeta}^{(k+1)}=\pi_r\left(\widetilde{\bbeta}^{(k+1)}\right)\end{aligned}$ \\
	}
	Record the solution as $\widehat{\bbeta}$.\\
	\Return $\widehat{\bbeta}$
\end{algorithm}

\par
\vspace{10pt}

The algorithmic convergence rate is presented in the following proposition.

\begin{proposition}
Let $\widehat{\bbeta}^{(k)}$ be the $k$ th iterated solution of Algorithm 1 . There exist constants $c_h$ and $C$, independent of $\left(n, p, s_0\right)$, such that when $h<c_h$, there exists $k<C \epsilon^{-2}$ and subgradient $u((\hat{\beta}^{(k)})^j) \in \partial|(\hat{\beta}^{(k)})^j|$, such that
$$
\left\|\nabla \widehat{R}_n\left(\widehat{\bbeta}^{(k)}\right)+\lambda \sum_{j=1}^p u\left((\hat{\beta}^{(k)})^j\right)\right\|_2 \leq \epsilon,
$$
where $\partial|\beta^j|$ denotes the sub-differential of the $\ell_1$-penalty function.
\end{proposition}

Proposition 1 justifies the rate at which the proposed algorithm converges. Specifically, the algorithm always converges to an approximate stationary solution  (known as an $\epsilon$-stationary solution) with a finite sample size. This means that after a certain number of iterations ($O\left(1 / \epsilon^2\right)$), the $\ell_2$ norm of the subgradient of the objective function is bounded by $\epsilon$ when the sample size is finite. When $k$ increases, the proposed algorithm will find the stationary solution that satisfies the subgradient optimality condition as $\epsilon \rightarrow 0$.

\subsection{Asymptotic analysis}\par
To establish the asymptotic normality, we need the following additional conditions. \par

\textbf{Assumption 2.}\par
(a) Assume $\widehat{\mathcal{M}}_{j_0}^{(n)}$ satisfies $\operatorname{Pr}\left(\left|\widehat{\mathcal{M}}_{j_0}^{(n)}\right| \leq \kappa_n\right)=1$ for some $1 \leq \kappa_n=o(n)$. Besides, there exists some constant $\alpha_0>1$ such that
$$
\operatorname{Pr}\left(\mathcal{M}_{j_0} \subseteq \widehat{\mathcal{M}}_{j_0}^{(n)}\right) \geq 1-O\left(\frac{1}{n^{\alpha_0}}\right).
$$\par

(b) Assume (i) $\operatorname{Pr}(\|\widetilde{\boldsymbol{\beta}}-\boldsymbol{\beta}_0\|_2 \leq \eta_n) \rightarrow 1$ for some $\eta_n>0$; (ii) $\eta_n \sqrt{\kappa_n \log p}=o(1)$ and $\sqrt{n} \eta_n^2=o(1)$; (iii) $\operatorname{Pr}(\|\widetilde{\boldsymbol{\beta}}_{\mathcal{M}_0^c}-\boldsymbol{\beta}_{0, \mathcal{M}_0^c}\|_1 \leq k_0\|\widetilde{\boldsymbol{\beta}}_{\mathcal{M}_0}-\boldsymbol{\beta}_{0, \mathcal{M}_0}\|_1) \rightarrow 1$ for some constant $k_0>0$. \par

 {(c) Assume there exists some constant $c_0>0$ such that $\left\|\boldsymbol{X}^T \boldsymbol{a}\right\|_{\psi_2} \leq c_0\|\boldsymbol{a}\|_2$ for any $\boldsymbol{a} \in \mathbb{R}^p$.} \par

\vspace{10pt}

 {Condition (a) requires that the feature screening procedure satisfies the sure screening property. Condition (b)  assumes that the initial estimator 
$\bbetat$ satisfies certain 
$\ell_1$ and $\ell_2$
  error bounds, which are crucial for ensuring valid subsequent inference. Condition (c) assumes that any linear combination of the covariates 
$\bX$ follows a sub-Gaussian distribution.}
Conditions (a)-(c) are also considered in \cite{shi}. We do not impose any stringer conditions on the design matrix. Also note that any stationary point of the program \eqref{ini}  satisfies the error bounds in (b) by Theorem 1.

\begin{theorem}
   {Assume  Assumption 1 and 2} hold. Assume $s_n \rightarrow \infty, s_n=o(n), \kappa_n^{5 / 2} \log p=$ $O\left(n / \log ^2 n\right)$ and $\kappa_n^3=O(n)$. Then, for any fixed $l \geq 1$, we have
$$
\sqrt{n} \Gamma_n^{*,(l-1)}(\hat{\beta}_{j_0}^{(l)}-\beta_{0, j_0}) \stackrel{d}{\rightarrow} N(0,1) .
$$  
\end{theorem}

Theorem 2 proves the validity of the proposed two-sided CI, for any $l \geq 1$. When $l=1$, $\hat{\beta}_{j_0}^{(l)}$ corresponds to the solution of the first-order approximation of the score equation. We note that \cite{Bickle}  and Ning and Liu (2017) used a similar one-step approximation to ensure the consistency of the resulting estimator. In practice, we can update $\hat{\beta}_{j_0}^{(l)}$ for a few Newton steps. In our numerical experiments, we find that $\hat{\beta}_{j_0}^{(l)}$ converges pretty fast and  $l$ is set as 8.


\section{Simulation studies}

This section carefully examines the finite-sample performance of our proposed method in simulation studies. To this end, we consider robust linear regression models consisting of contaminated models and heavy-tailed error models, in which nonconvex loss (Tukey’s biweight loss) and convex loss (Pseudo Huber loss) are implemented respectively to demonstrate that our proposed method can accommodate both types of loss functions. In each table, we report the empirical probability (ECP) and average length (AL) of the CIs.

\subsection{Contaminated models}
We generated random covariates $\bX_i \sim \mathrm{N}\left(\mathbf{0}, \boldsymbol{\Sigma}\right)$, where $
\boldsymbol{\Sigma}=\left\{0.5^{|i-j|}\right\}_{i, j=1, \ldots, p}
$ and responses $Y_i=\boldsymbol{X}_i^T \boldsymbol{\beta}_0+\varepsilon_i,$ where $\beta_{0,1}=3, \beta_{0,2}=1.5, \beta_{0,3}=0,  \beta_{0,4}=0,  \beta_{0,5}=2$ and $ \beta_{0,j}=0$ for $j>5.$  Here  we study contaminated models for the noise, namely $\varepsilon_i \sim 0.9 \mathrm{N}(0,1)+0.1\mathrm{N}\left(0, \sigma^2\right)$. For the loss function, we use Tukey’s loss  with $t_0 = 4.685$. 

We consider the following two settings: \ 
(A) $\sigma=5.$\ \  (B) $\sigma=10.$ 
In both settings, we set $n=500, p=1000$. We aim to construct two-sided CIs for $\beta_{0,1}, \beta_{0,2}, \beta_{0,3}, \beta_{0,5}$.

 To calculate the CIs, we set $s_n=\lfloor 2 n / \log (n)\rfloor$ and $l=8.$ We estimate $\mathcal{M}_{j_0}$ by SIRS \citep{SIRS}.  The adaptive LASSO estimator is used as the initial estimator $\widetilde{\boldsymbol{\beta}}$.  {For the penalization parameter 
$\lambda$, we recommend using cross-validation to select the optimal value. Specifically, one can minimize a cross-validation score, 
CV($\lambda$), which can be  a measure of  model fitting or prediction error, over a finite grid of candidate 
$\lambda$ values.}
We use projected gradient descent to solve the system with $r = 10$. 

Comparison is made with CIs constructed by linear regression with least square loss in ROSE and Debiased LASSO. Results are averaged over 500 simulations.

\begin{table}[H]

\caption{ECP and AL of the CIs of contaminated models with standard errors in parenthesis}

\centering
\begin{tabular}{cccccccc}
\hline \multicolumn{2}{c}{ Settings } & \multicolumn{2}{c}{ Robust-ROSE } & \multicolumn{2}{c}{ ROSE-Linear } & \multicolumn{2}{c}{ DLASSO } \\ \hline
\multicolumn{2}{c}{  } & ECP(\%)& AL*100&ECP(\%)& AL*100&ECP(\%)& AL*100 \\
\hline \multirow{4}{*}{$\sigma=5$} & $\beta_1$ & $94.6$ & $21.5(1.96)$ & $92.8$ & $33.7(3.29)$ & 0.0  & 35.4(3.69) \\
& $\beta_2$ & $96.4$ & $21.3(1.79)$ & $92.2$  & $34.1(3.36)$ & 0.2 & 35.4(3.63)\\
& $\beta_3$ & $96.8$ & $21.5(1.99)$ &$91.0$ & $34.3(3.41)$&100.0& 35.4(3.67)\\
& $\beta_5$ & $95.4$ & $21.5(1.88)$ & $93.8$ &$34.0(3.33)$&0.0&35.5(3.64)\\
\hline 
\multirow{4}{*}{$\sigma=10 $} & $\beta_1$ & $95.2$ & $21.1(1.87)$ &$90.4$ & $59.8(7.50)$& 0.0 & 64.3(8.01) \\
& $\beta_2$ & $94.2$ & $20.8(1.73)$ & $91.8$ &$60.7(7.42)$ & 0.0 &63.3(7.88) \\
& $\beta_3$ & $95.6$ & $21.1(1.88)$ & $92.4$&$61.3(7.51)$& 100.0 & 64.3(7.95) \\
& $\beta_5$ & $95.2$ & $21.1(1.89)$ & $90.0$&$60.6( 7.37)$& 0.0 & 64.4(7.89) \\
\hline
\end{tabular}

 \label{tab:1}   
\end{table}

Table \ref{tab:1} illustrates that the ECPs of our method closely align with the nominal level for all values of $\beta$ and $\sigma$. However,
the ECPs of ROSE are lower than the nominal level, and as $\sigma$ increases, indicating greater contamination, the deviation from the nominal level also increases for all values of $\beta$. Besides, CIs produced by DLASSO consistently fail to fully encompass the true values for non-zero $\beta$'s, but consistently cover zero. As a result, they exhibit diminished coverage power overall.

Moreover, our procedure achieves the shortest average length (AL) among all the methods, thereby surpassing the other two methods in terms of both ECP and AL.

\subsection{Heavy-tailed error models}

Now, we consider the heavy-tailed error case. We generated random covariates $\bX_i \sim \mathrm{N}\left(\mathbf{0}, \boldsymbol{\Sigma}\right)$, where $
\boldsymbol{\Sigma}=\left\{0.5^{|i-j|}\right\}_{i, j=1, \ldots, p}
$  and responses $Y_i=\boldsymbol{X}_i^T \boldsymbol{\beta}_0+\varepsilon_i,$ where $\beta_{0,1}=1, \beta_{0,2}=1, \beta_{0,3}=1$ and $ \beta_{0,j}=0$ for $j>3.$  The error $\varepsilon_i = (-1)^{V_i}\cdot U_i,$ where $U_i \sim \text{Lognormal} (0,1)$ and $V_i = 1 $ or 2 with equal probability. For the loss function, we use pseudo Huber loss. 

We set $n=500, p= 1000$. The model selection and other configurations are set the same as in the contaminated models scenario. The two alternative methods under comparison are also identical. Results are averaged over 400 simulations.

\begin{table}[H]
\caption{ECP and AL of the CIs of heavy-tailed error models with standard errors in parenthesis}

\centering
\setlength{\abovecaptionskip}{10pt}

\begin{tabular}{cccccccc}
\hline \multicolumn{2}{c}{ Settings } & \multicolumn{2}{c}{ Robust-ROSE } & \multicolumn{2}{c}{ ROSE-Linear } & \multicolumn{2}{c}{ DLASSO } \\ \hline
\multicolumn{2}{c}{  } & ECP(\%)& AL*100&ECP(\%)& AL*100&ECP(\%)& AL*100 \\
\hline \multicolumn{2}{c}{$\beta_1$} & $94.75$ & $32.7(2.80)$ & $91.5$ & $49.9(6.38)$ & 0.25  & 52.3(7.88) \\
\multicolumn{2}{c}{$\beta_2$} & $94.25$ & $32.8(2.93)$ & $95.0$  & $50.1(6.52)$ & 0.0 & 52.4(8.00)\\
\multicolumn{2}{c}{$\beta_3$} & $94.75$ & $32.9(2.91)$ &$91.5$ & $50.1(6.75)$&0.0& 52.5(7.87)\\
\multicolumn{2}{c}{$\beta_5$} & $95.25$ & $32.5(2.69)$ & $90.75$ &$50.5(6.56)$&100.0&52.5(8.02)\\
\hline 

\end{tabular}

    \label{tab:2}
\end{table}

It can be seen from Table \ref{tab:2} that our method still yields the shortest CIs while maintaining  ECPs that are very  close to the nominal level. The coverage efficiency of CIs by ROSE is still impacted by the heavy-tailed behavior of errors, and DLASSO exhibits a similar pattern.

\section{Real data analysis}

\par
We will now turn our attention to the examination of the real dataset of riboflavin (vitamin B2) production in Bacillus subtilis. We propose to use robust linear regression to identify significant genes. We center the response and standardize all the covariates before analysis and then construct CIs for each individual coefficient and apply Bonferroni’s method for multiple adjustments. At the $5\%$ significance level, our method finds seven important genes (the 1303rd, 1516th, 3311st, 4002nd,
4003rd, 4004th and 4006th).  { The 4003rd and 4004th genes coincide with those found in \cite{Java} and the 1303rd and 4003rd were also detected in \cite{Hila}}.

  \begin{figure}[H]
    \centering
    \includegraphics[width=7cm]{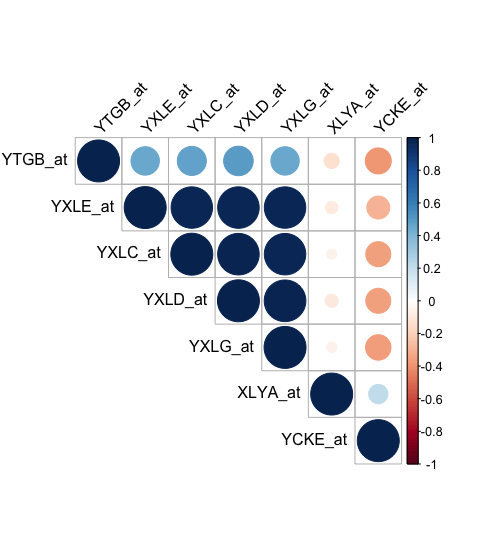}
    \caption{The Correlogram of the detected genes}
   \label{Ch3-figure: corrplot}
\end{figure}

We investigated the pairwise correlation among the identified genes, and a correlogram is presented below. It is evident that genes 4002nd, 4003rd, 4004th, and 4006th exhibit high correlation in each pair. This observation is further supported by existing biological evidence, indicating that these genes collectively belong to the SigY regulon  {\citep{subtiwiki}}.

In contrast, \cite{shi} only claimed three important genes (the 1588th, 3154th, and 4004th). Moreover, the de-sparsified Lasso approach claims no variables are significant.

\section{Conclusion}

In this paper, we developed a recursive online score estimation (ROSE) methodology tailored for high-dimensional robust linear regression models, extending the ROSE framework of Shi et al. (2021) to more general and challenging scenarios. We thoroughly analyzed the behavior of stationary points in the nonconvex M-estimation landscape and proposed a practical algorithm with robust computational guarantees. Building on an initial estimator, we introduced a robust ROSE estimation procedure and established its asymptotic normality, enabling the construction of confidence intervals for parameters of interest. Our simulation studies and real-world data analysis highlight the necessity and advantages of our approach compared to existing methods. The results demonstrate the effectiveness of our methodology in addressing the dual challenges of robustness and high dimensionality, providing a significant contribution to the field of high-dimensional statistical inference.

\bibliographystyle{apalike}
\bibliography{ref}

\end{document}